\documentclass{article}
  \usepackage{emulateapj}
  \usepackage{psfig}

\lefthead{Brandner et al.}
\righthead{HST/NICMOS detection of a pre-main-sequence
population in the 30 Doradus Nebula}

\begin{document}

\title{HST/NICMOS detection of a partially embedded, 
intermediate-mass 
pre-main-sequence population in the 30 Doradus Nebula}\footnote{Based on 
observations 
with the NASA/ESA Hubble Space Telescope obtained at the Space 
Telescope Science Institute, which is operated by the Association of 
Universities for Research in Astronomy, Inc., under the NASA contract 
NAS5-26555, and observations at the European Southern Observatory, La Silla, 
obtained during technical time.}

\author{Wolfgang Brandner\altaffilmark{2,3}, 
Eva K.\ Grebel\altaffilmark{4},
Rodolfo H.\ Barb\'a\altaffilmark{5,6},
Nolan R.\ Walborn\altaffilmark{7},
Andrea Moneti\altaffilmark{8}}

\affil{$^2$University of Hawaii, Institute for Astronomy, 2680 Woodlawn Dr., 
Honolulu, HI 96822, USA}
\affil{$^3$European Southern Observatory, Karl-Schwarzschild-Str.\ 2, D-85748
Garching, Germany}
\authoremail{wbrandne@eso.org}
\affil{$^4$Max-Planck-Institut f\"ur Astronomie, Am K\"onigstuhl 17, D-69117 
Heidelberg, Germany}
\authoremail{grebel@mpia-hd.mpg.de}
\affil{$^5$Facultad de Ciencias Astron\'omicas y Geof\'{\i}sicas, Universidad 
Nacional de la Plata, Paseo del Bosque S/N, 1900 La Plata, Argentina}
\affil{$^6$Member of the Carrera del Investigador Cient\'{\i}fico y 
Tecnol\'ogico, CONICET, Argentina}
\authoremail{rbarba@fcaglp.fcaglp.unlp.edu.ar}
\affil{$^7$Space Telescope Science Institute, 3700 San Martin Drive, Baltimore,
MD 21218, USA}
\authoremail{walborn@stsci.edu}
\affil{$^8$Institut d'Astrophysique Paris, 98bis Blvd Arago, F-75014 Paris, France}
\authoremail{moneti@iap.fr}

\begin{abstract} 
We present the detection of an intermediate-mass pre-main-sequence population
embedded in the nebular filaments surrounding the 30 Doradus region 
in the Large Magellanic Cloud (LMC) using HST/NICMOS.
In addition to four previously known luminous Class I infrared ``protostars,''
the NICMOS data reveal 20 new sources with intrinsic infrared
excess similar to Galactic pre-main sequence stars. Based on their
infrared brightness, these objects can be identified as the LMC equivalent of
Galactic pre-main sequence stars. The faintest LMC Young Stellar Objects
in the sample have colors similar to T Tauri and have about the same 
brightness as T Tauri if placed at the distance of the LMC.
We find no evidence for a lower-mass cut-off in the initial mass function.
Instead, the
whole spectrum of stellar masses from pre-main sequence stars with 
$\sim$1.5\,M$_\odot$ 
to massive O stars still embedded in dense knots appears to be present in
the nebular filaments. 
The majority of the young stellar objects can be found to the north
of the central starburst cluster R136. This region is very likely 
evolving into an OB association.

The observations provide further evidence that star formation in the
30 Doradus region is very similar to Galactic star formation, and
confirm the presence of sequential star formation
in 30 Doradus, with present-day star formation taking place in  
the arc of molecular gas to the north and west of the starburst cluster.

\end{abstract}


\section{Introduction}

The study of extragalactic star formation began 20 years ago
with the identification of the first candidate protostar in N159 in the
Large Magellanic Cloud (LMC) by Gatley et al.\ (1981). It was
subsequently expanded by the identification of other candidate
``protostars'', in particular in the 30 Doradus region (Hyland et al.\ 1992).
30 Doradus is the most luminous giant HII region in the Local Group 
(Kennicutt 1984). The starburst region consists of multiple stellar 
generations ranging from 2\,Myr old Wolf-Rayet and O3 main-sequence
stars to evolved blue and red supergiants with ages up to 25\,Myr 
(Walborn \& Blades 1997; de Koter et al.\ 1998; Massey \& Hunter 1998;
Grebel \& Chu 2000; and references therein). The stellar population has 
been studied
with high spatial resolution with the Hubble Space Telescope (HST) in the 
optical down to 2.8\,M$_\odot$ (Hunter et al.\ 1995) and with adaptive optics 
in the near-infrared (Brandl et al.\ 1996). Recently,
Sirianni et al.\ (2000) extended the study of the initial
mass function down to 1.35\,M$_\odot$.  Infrared and radio observations
reveal that molecular gas and warm dust is concentrated in an arc to the north
and west of the central starburst cluster in 30\,Doradus (Werner et al.\ 1978;
Johansson et al.\ 1998; Rubio et al.\ 1998). 
Three early O-type stars embedded in dense nebular knots (Walborn \& Blades 
1987, 1997), four luminous infrared ``protostars'' (Hyland et al.\ 1992) as well as
numerous infrared sources associated with ongoing star formation (Rubio et 
al.\ 1992, 1998) have been identified in the arc.
Using HST/NICMOS (Thompson et al.\ 1998), we studied the embedded IR sources 
in the nebular filaments
and their environment in a multi-color survey, aiming at a better 
understanding of the extent of triggered star formation, the nature of the 
embedded stellar population, and its luminosity function (see Walborn et al.\ 
1999 for first results). The results of a complementary HST/NICMOS
program, which aimed at establishing a deep H-band luminosity function for
stars in the central starburst cluster, are reported elsewhere
(Zinnecker et al.\ 1999, and in prep.).

Section 2 gives an overview on the observations. Section 3 details the
identification of pre-main sequence stars. In Section 4, the new results 
are discussed with respect to previous studies of young stellar objects
in the LMC, and Section 5 presents a brief summary.

\section{Observations and data reduction} 

\begin{figure*}[htb]
\centerline{
\psfig{figure=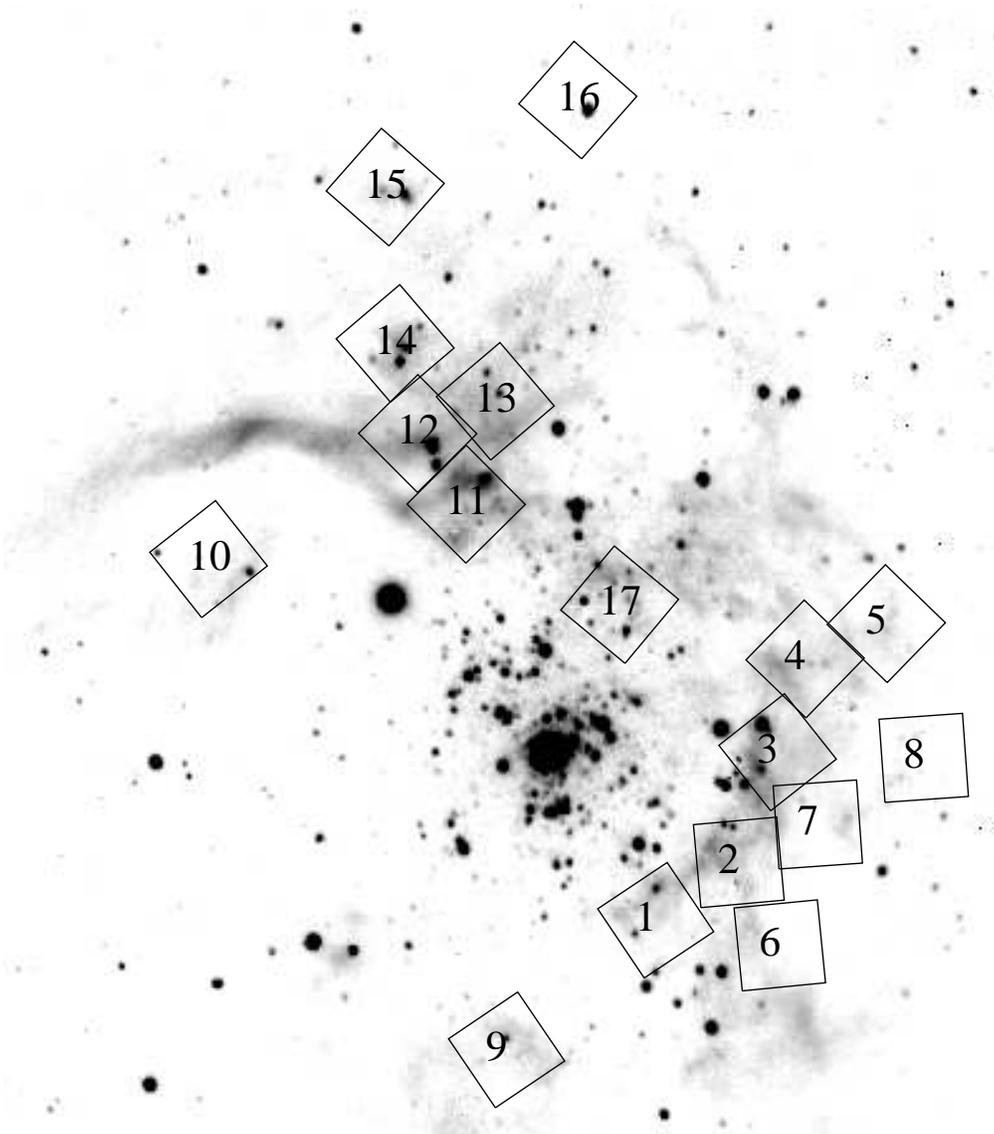,angle=0,width=14.0cm}
}
\figcaption[brandner.fig1.ps]{200$''$ $\times$ 225$''$ K-band image (obtained with IRAC-2a at
the MPI/ESO 2.2m telescope) of the 30 Doradus region
indicating the location and orientation of the HST/NICMOS (NIC2) pointings.
North is up and East is to the left.
\label{fig1}}
\end{figure*}

\subsection{Ground-based near-infrared imaging}

Ground-based near-infrared imaging observations in the J-, H-, and K-band
were carried out on 1993 February 15 with the ESO/MPI 2.2m telescope
and the IRAC-2a camera (Moorwood et al.\ 1992) at La Silla, Chile. 
IRAC-2a was equipped with
a NICMOS3 array, and a pixel scale of 0\farcs49 pixel$^{-1}$ was chosen.
The data were obtained as a 3$\times$3 mosaic covering an area of
approximately 350$''$ $\times$ 350$''$. Integration
times were 60s per position and filter, and  the seeing was 1\farcs0 to 
1\farcs2.

The data were reduced in the standard way, used to identify embedded red
objects, and --- combined with the data obtained by Rubio et al.\ (1998) ---
helped to plan the HST/NICMOS observations. Figure \ref{fig1} shows
a 200$''$ $\times$ 225$''$ subset of the K-band mosaic centered
on 30 Doradus.

\subsection{Space-based near-infrared imaging}

\begin{table*}[htb]
\caption{Candidate Class I sources and Herbig AeBe/T\,Tauri stars to the south
and west of R136a\label{tab1}}
\begin{tabular}{llccccc} \tableline \tableline
Name&Alias&$\alpha$(2000) &$\delta$(2000) &J & J--H & H--K$_{\rm s}$\\
&               &[hms]      &[$^\circ$ $'$ $''$] & [mag] & [mag] & [mag]  \\ \tableline
30Dor-NIC08a&                      &5 38 29.87& -69 06 11.3 & 16.95$\pm$0.06 & 0.44$\pm$0.07 &0.55$\pm$0.14\\
30Dor-NIC07a&                      &5 38 33.14& -69 06 11.1 & 18.68$\pm$0.12 & 1.84$\pm$0.15 &1.75$\pm$0.11\\
30Dor-NIC07b&                      &5 38 33.78& -69 06 15.4 & 20.03$\pm$0.11 & 0.86$\pm$0.14 &0.79$\pm$0.14\\
30Dor-NIC03a&[HJ91] P3\tablenotemark{1} , IRSW-30\tablenotemark{2}     &5 38 34.57& -69 05 58.2 & 15.23$\pm$0.09 & 2.16$\pm$0.10 &2.05$\pm$0.05\\
30Dor-NIC02a&                      &5 38 36.72& -69 06 21.0 & 18.40$\pm$0.09 & 0.83$\pm$0.11 &0.78$\pm$0.08\\
30Dor-NIC01a&                      &5 38 40.12& -69 06 37.4 & 19.28$\pm$0.09 & 0.36$\pm$0.13 &0.65$\pm$0.17\\
30Dor-NIC01b&                      &5 38 40.21& -69 06 33.2 & 19.21$\pm$0.07 & 0.78$\pm$0.09 &0.67$\pm$0.06\\
30Dor-NIC09a&                      &5 38 43.47& -69 07 00.7 & 19.97$\pm$0.10 & 0.62$\pm$0.13 &0.58$\pm$0.13\\ \tableline
\end{tabular}

\tablerefs{$^1$ Hyland \& Jones (1991), $^2$ Rubio et al.\ (1998)}

\end{table*}

\begin{table*}[htb]
\caption{Candidate Class I sources and Herbig AeBe/T\,Tauri stars to the north
and north-east of R136a\label{tab2}}
\begin{tabular}{llccccc} \tableline \tableline
Name&Alias&$\alpha$(2000) &$\delta$(2000) &J & J--H & H--K$_{\rm s}$\\
&               &[hms]      &[$^\circ$ $'$ $''$] & [mag] & [mag] & [mag]  \\ \tableline
30Dor-NIC16a&[HJ91] P2$^1$         &5 38 41.62& -69 03 54.7 & 15.38$\pm$0.05 & 2.11$\pm$0.10 &1.56$\pm$0.12\\
30Dor-NIC13a&                      &5 38 44.65& -69 05 01.6 & 18.03$\pm$0.06 & 0.60$\pm$0.07 &0.57$\pm$0.05\\
30Dor-NIC13b&                      &5 38 45.00& -69 04 57.9 & 18.90$\pm$0.06 & 0.71$\pm$0.08 &1.13$\pm$0.06\\
30Dor-NIC13c&                      &5 38 45.10& -69 04 54.9 & 19.46$\pm$0.08 & 0.74$\pm$0.11 &0.79$\pm$0.09\\
30Dor-NIC11a&                      &5 38 45.96& -69 05 17.0 & 20.05$\pm$0.13 & 0.62$\pm$0.17 &0.74$\pm$0.14\\
30Dor-NIC11b&                      &5 38 46.57& -69 05 20.1 & 17.65$\pm$0.04 & 0.64$\pm$0.05 &1.22$\pm$0.04\\
30Dor-NIC12a&                      &5 38 46.99& -69 05 08.3 & 19.46$\pm$0.08 & 1.05$\pm$0.10 &1.55$\pm$0.07\\
30Dor-NIC12b&[HJ91] P1\tablenotemark{1} , IRSN-122\tablenotemark{2}         &5 38 47.11& -69 05 06.1 & 17.46$\pm$0.09 & 2.16$\pm$0.11 &2.09$\pm$0.08\\
30Dor-NIC12c&                      &5 38 47.24& -69 05 03.4 & 19.86$\pm$0.12 & 1.34$\pm$0.14 &1.09$\pm$0.12\\
30Dor-NIC12d&[HJ91] P1\tablenotemark{1} , IRSN-126\tablenotemark{2}         &5 38 47.25& -69 05 02.3 & 17.15$\pm$0.06 & 2.83$\pm$0.07 &2.54$\pm$0.04\\
30Dor-NIC15a&                      &5 38 48.30& -69 04 10.3 & 19.01$\pm$0.08 & 0.77$\pm$0.10 &0.67$\pm$0.09\\
30Dor-NIC15b&[HJ91] P4\tablenotemark{1} , IRSN-134\tablenotemark{2}         &5 38 48.45& -69 04 12.0 & 17.06$\pm$0.07 & 1.82$\pm$0.08 &1.76$\pm$0.04\\
30Dor-NIC14a&                      &5 38 48.76& -69 04 47.6 & 19.16$\pm$0.09 & 0.61$\pm$0.11 &0.64$\pm$0.12\\
30Dor-NIC14b&                      &5 38 48.82& -69 04 47.1 & 19.91$\pm$0.12 & 0.92$\pm$0.16 &0.97$\pm$0.17\\
30Dor-NIC12e&                      &5 38 48.87& -69 05 02.2 & 20.30$\pm$0.14 & 0.79$\pm$0.18 &0.89$\pm$0.16\\
30Dor-NIC10a&                      &5 38 55.55& -69 05 25.8 & 16.35$\pm$0.03 & 0.36$\pm$0.04 &0.41$\pm$0.03\\ \tableline
\end{tabular}

\tablerefs{$^1$ Hyland \& Jones (1991), $^2$ Rubio et al.\ (1998)}

\end{table*}

Observations of 17 fields centered on infrared sources and one background field
(``sky'') were obtained with HST/NICMOS (NIC2) in Feb.\ and Mar.\ 1998.
14 of the fields were observed in the filters F110W (320s), 
F160W (384s), and F205W (448s). Field 17 was only observed in selected 
narrow-band filters, i.e., F187N (Pa$\alpha$), F212N (H$_2$), F215N
(cont.), and F216N (Br$\gamma$), while observations of fields 4 and 5
were only partially executed due to instrument problems (NICMOS suspended
due to a particle hit).
The location and orientation
of the individual NIC2 fields are outlined in Figure \ref{fig1}.
In the present paper, we focus on the photometric analysis of the
broad-band data.

The data were reduced using synthetic  darks customized for the appropriate
detector temperature. The observations of the background field in F205W were 
combined to produce a sky frame. This ``sky'' was subtracted
from the on-source observation in F205W in order to remove the thermal
emission of the telescope. Source identification and
photometry were carried out with the IRAF implementation of DAOPHOT.
The uncertainties quoted in Tables \ref{tab1} to \ref{tab3} are
DAOPHOT fitting uncertainties.
Transformations from the HST system to
ground-based photometry were computed by a comparison of HST/NICMOS
observations of stars from the NICMOS list of stars for photometric
transformations (NICMOS Data Handbook Version 4)
to photometry of the same stars as presented in the
2MASS point source catalog (see also Skrutskie et al.\ 1997; 
Beichman et al.\ 1998).
We derived the following color transformations (in units of [mag]):
$${\rm m}_{\rm 110} - {\rm J}  = (0.263\pm0.020) * ({\rm m}_{\rm 110} - {\rm m}_{\rm 160}) + (0.083\pm0.022)$$
$${\rm m}_{\rm 160} - {\rm H}  = (0.072\pm0.041) * ({\rm m}_{\rm 110} - {\rm m}_{\rm 160}) + (0.032\pm0.045)$$
$${\rm m}_{\rm 205} - {\rm K_s}  = (0.268\pm0.063) * ({\rm m}_{\rm 160} - {\rm m}_{\rm 205}) - (0.038\pm0.070)$$

The color transformations are derived for standard stars with 
J--H colors ranging from $-$0.2$^{\rm m}$ to $+$2.0$^{\rm m}$ and
H--K$_{\rm s}$ colors ranging from $+$0.1$^{\rm m}$ to $+$0.9$^{\rm m}$.
The color transformations from F110W and F160W to ground-based J and H
magnitudes are within the uncertainties in good agreement with the
color transformations presented by Origlia \& Leitherer (2000).
Transformations from the HST NICMOS system to the CTI/CTIO photometric system,
and possible complications due to spectral features which are not in common
between standard stars and program stars, are discussed by Stephens 
et al.\ (2000). 

We estimate the absolute photometric uncertainty to be of the
order of $\pm$0\fm1. The brightness limit for sources detected
in all three broad-band filters is m$_{\rm K}$ $\approx$ 19\fm8 for
a signal-to-noise ratio of $\approx$7. Relative photometric uncertainties
for individual sources are quoted in Tables \ref{tab1} to \ref{tab3}.

Positions of individual sources are based on the astrometric
information in the FITS file headers. The relative astrometric uncertainty
in the position of individual sources within one NICMOS frame should be of 
the order of $\pm$0\farcs02,
whereas the absolute positional uncertainty might be as large as  $\pm$1$''$.

\section{Identification of pre-main sequence stars}

\begin{figure*}[ht]
\vbox{
\hbox{
\psfig{figure=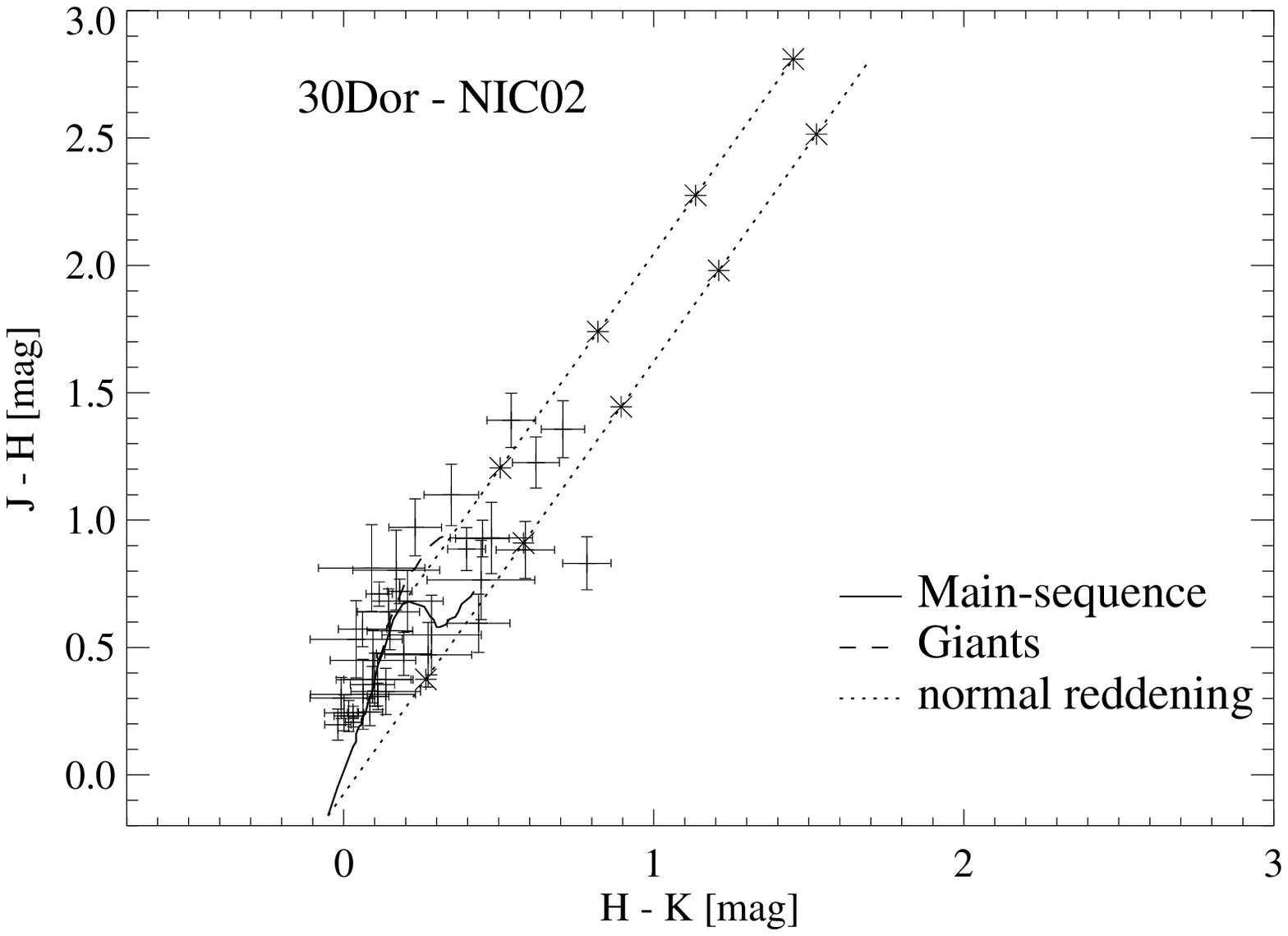,angle=0,width=8.4cm}
\psfig{figure=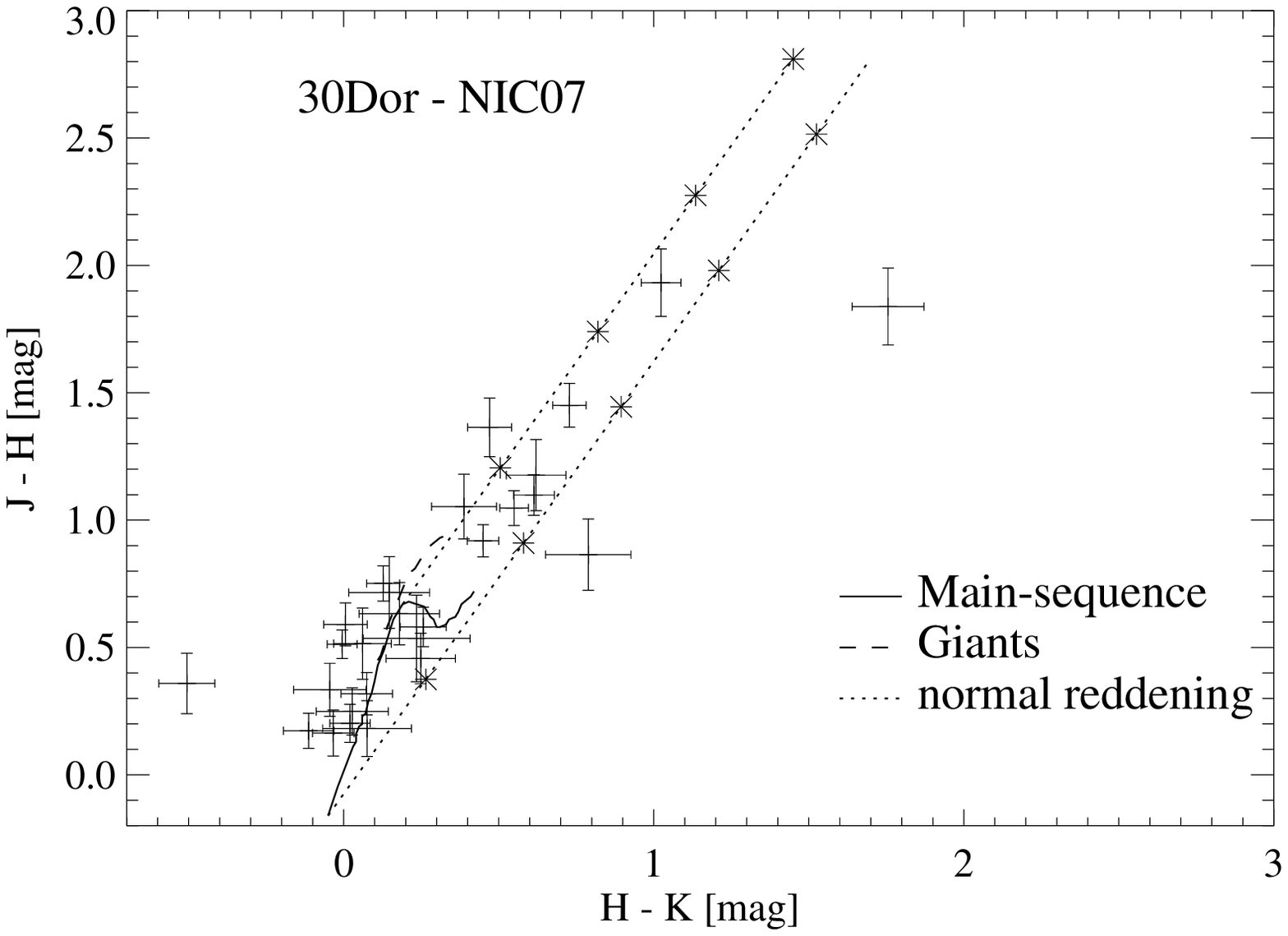,angle=0,width=8.4cm}
}
\hbox{
\psfig{figure=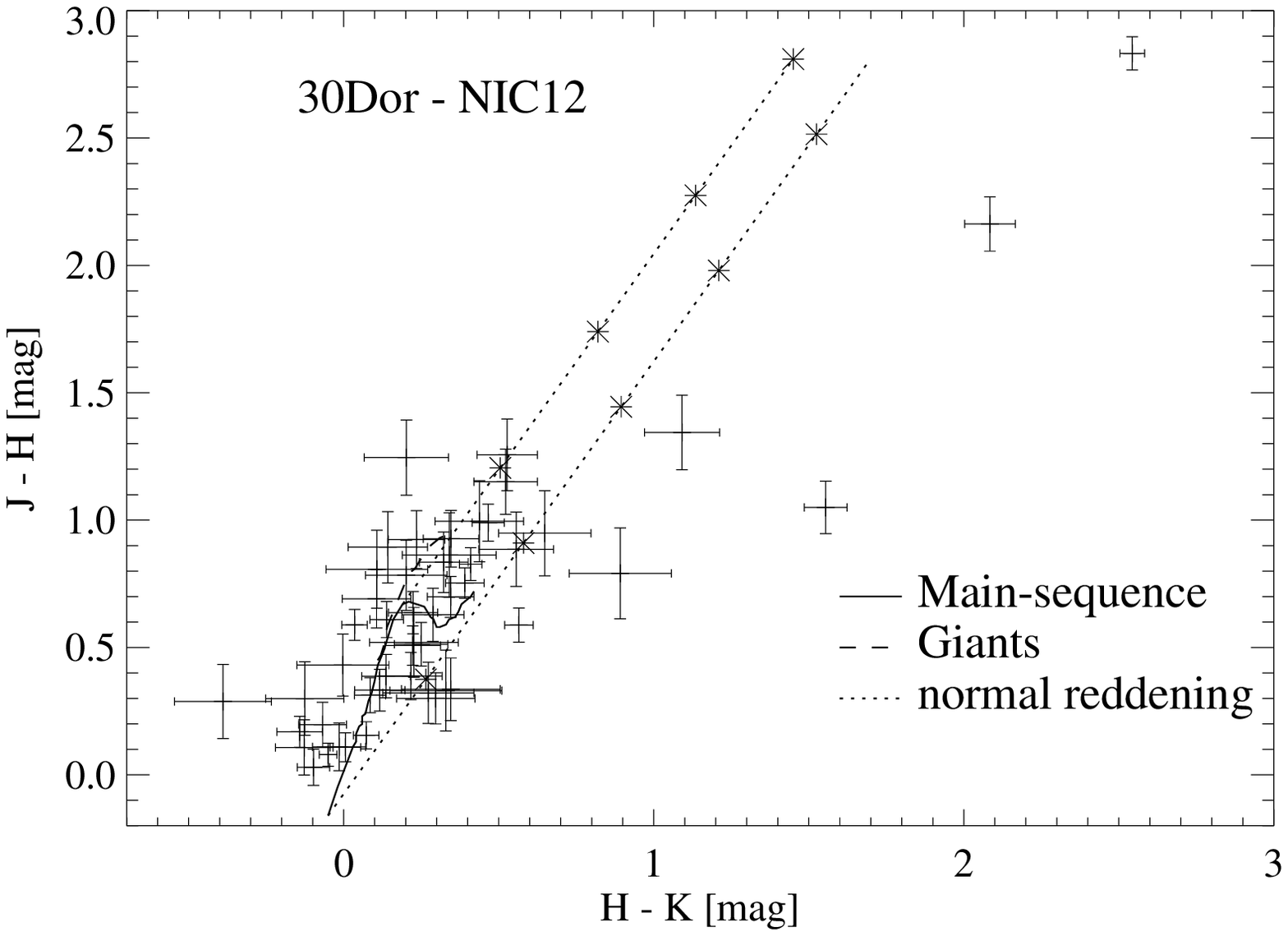,angle=0,width=8.4cm}
\psfig{figure=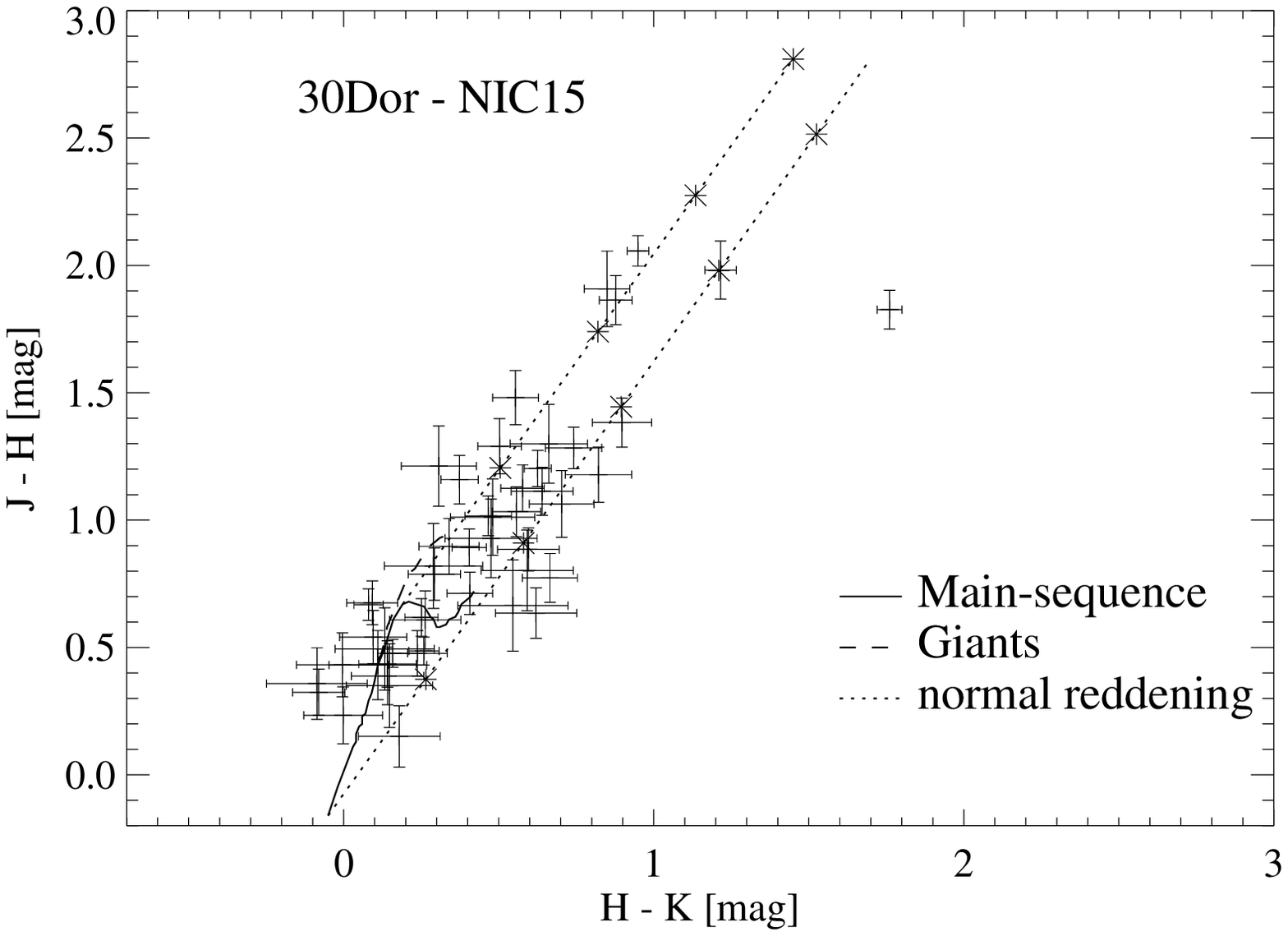,angle=0,width=8.4cm}}
}
\figcaption[brandner.fig2a.ps,brandner.fig2b.ps,brandner.fig2c.ps,brandner.fig2d.ps]{Examples of near-infrared color-color diagrams of NIC2 fields
to the west (top)
and to the north of R136a (bottom). The location of (unreddened) main-sequence
stars is indicated by a solid line, the location of giant stars by
a dashed line. The dotted lines indicate the
reddening band for normal extinction. Asterisks mark steps of 5$^{\rm m}$
in visual extinction.
Stars with intrinsic infrared excess fall outside the region of
normal reddening. In addition to previously known luminous Class I
sources, stars with the color characteristics of intermediate- to low-mass
young stellar objects (Herbig AeBe and T\,Tauri stars) are
detected.\label{fig2}}
\end{figure*}

Near infrared (NIR) color-color diagrams can be used to separate
stars with infrared excess due to extinction from stars with intrinsic
infrared excess. Intrinsic infrared excess can be attributed to the presence
of circumstellar material, and thus is a sign of youth.
As pointed out by Lada \& Adams (1992), different types of objects
occupy well-defined, distinct
regions in NIR color-color diagrams.
Many young stellar objects lie outside the normal reddening band.
Infrared ``protostars'' (Class I sources) typically are highly reddened
(A$_{\rm V}$ = 10$^{\rm m}$ to 40$^{\rm m}$) and have both J--H and
H--K colors $\ge$ +1\fm5. Herbig AeBe stars tend to have J--H and
H--K colors between +0\fm8 and +1\fm5. Classical (CTTS) and weak-line T Tauri 
stars (WTTS) occupy a region characterized by lower extinction 
(A$_{\rm V}$ $\le$ 10$^{\rm m}$). CTTS
cluster around J--H $\approx$ +1\fm0 and H--K $\approx$ +0\fm8, whereas WTTS
typically fall within the reddening band.
Classical Be stars on the other hand can be found near the blue end
of the main-sequence. They show less intrinsic reddening than Herbig AeBe
or T\,Tauri stars and tend to have J--H colors
between $-$0\fm2 and +0\fm3 and H--K colors between $-$0\fm4 and +0\fm3.

Variability of individual sources and other effects like, e.g.,
source geometry make it difficult to determine the physical properties
of individual young stellar objects. 
The evolutionary nature of a population of young stellar objects can,
however, be deduced on a statistical basis from their
location in a J--H vs.\ H--K color-color diagram.
Tables \ref{tab1} and \ref{tab2} list all sources which were detected
in all three bands and which exhibit intrinsic IR excess
(i.e., they lie within the photometric uncertainties clearly outside
the standard reddening band).

\begin{figure*}[htb]
\psfig{figure=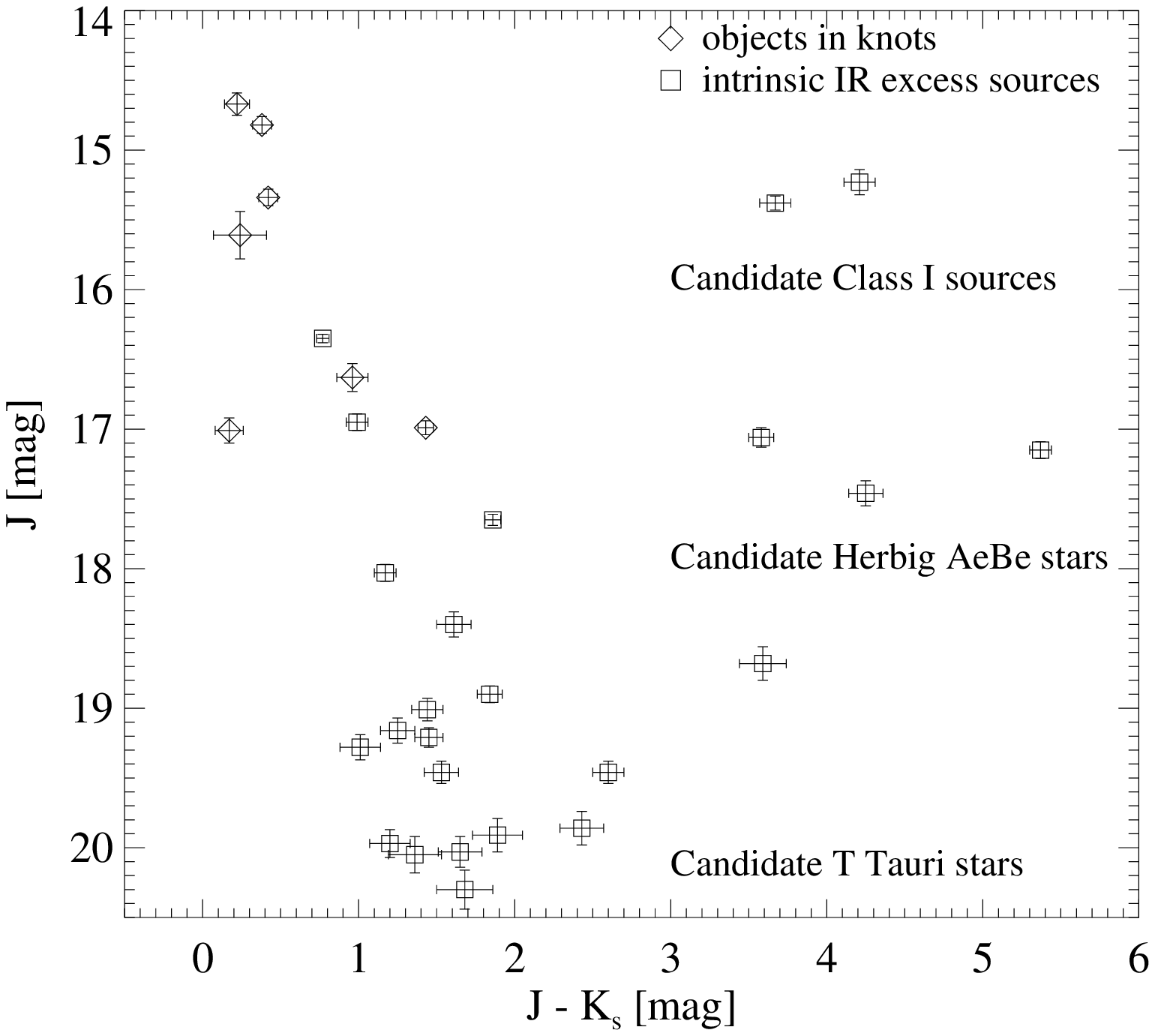,angle=0,width=14.0cm}
\figcaption[brandner.fig3.ps]{J vs.\ J-K$_{\rm s}$ color-magnitude diagram for candidate
Young Stellar Objects and the brightest objects in Knot 1 to 3
(see Tables 1 to 3). The approximate J-band brightness for
subsamples corresponding to candidate Class I sources, Herbig AeBe stars
and T Tauri stars in the LMC is indicated.
\label{fig3}}
\end{figure*}

\begin{figure*}[htb]
\centerline{
\psfig{figure=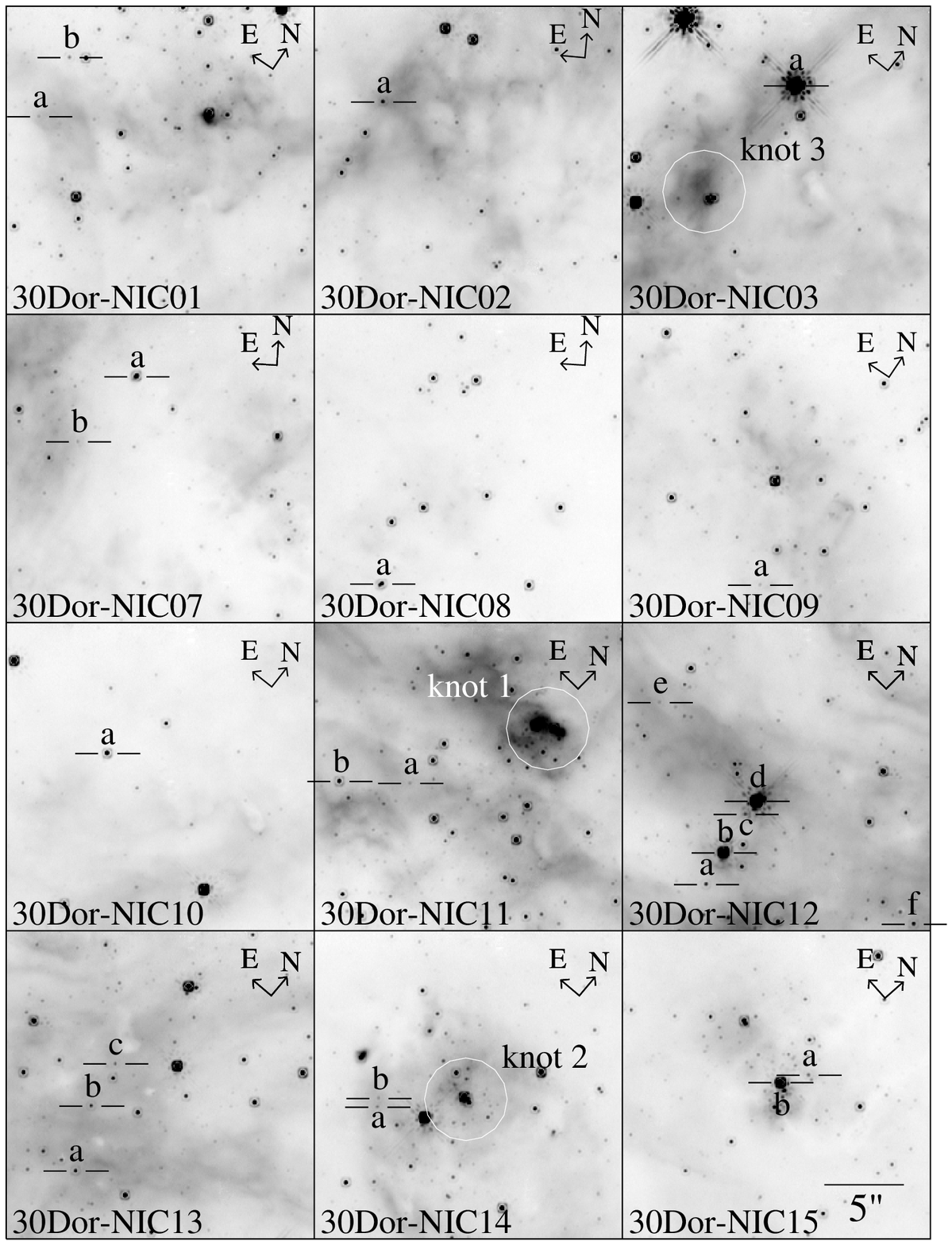,angle=0,width=15.0cm}}
\figcaption[brandner.fig4.ps]{Finding charts for pre-main sequence candidates in 12 of the
NICMOS/NIC2 fields in 30 Doradus. The field of view of each individual
field is 19$''$ $\times$ 19$''$, and the images shown have been observed
through the F205W filter. Note that 30Dor-NIC12f is identical to
30Dor-NIC13a. The locations of knot 1 to 3 (Walborn \& Blades 1987, 1997) are
indicated by circles.\label{fig4}}
\end{figure*}

Figure \ref{fig2}, top, shows NIR color-color diagrams of
two fields centered approximately 30$''$ and 50$''$ to the west of R136a.
In addition to stars with intrinsic colors close to the colors of
unreddened dwarf and giant stars, some stars show clear evidence of IR 
excess. The majority of the stars with IR excess are located in the
area of the normal reddening band, and their IR excess is very
likely due to enhanced (foreground) extinction of A$_{\rm V}$ = 5$^{\rm m}$ to
10$^{\rm m}$.

Figure \ref{fig2}, bottom,
shows the color-color diagrams for two of the fields located
to the north of R136a. These fields (30Dor-NIC 11, 12, 13, 14, and 15)
house a relatively large number of stars
with intrinsic NIR excess. Based on their
locations in the NIR color-color diagram, the objects appear to be
Class I ``protostars'' and Herbig AeBe stars or T\,Tauri stars 
(see Lada \& Adams 1992).
Figure \ref{fig3} shows a color-magnitude diagram for the objects listed
in Tables 1 to 3. The faintest stars in the sample have J$\approx$20\fm0.
According to the 2MASS point source catalog (Second Incremental
Data Release, Cutri et al.\ 2000), T\,Tauri itself has
J=7\fm26, J--H=1\fm02, and H--K=0\fm91.
If moved from its location in the Taurus T association (distance 140 pc)
to the distance of the LMC (distance $\approx$ 50 kpc), T\,Tauri would
have J=20\fm0. Thus the faintest objects listed in Table \ref{tab1} and
\ref{tab2} could actually be LMC counterparts to T\,Tauri.

The other fields to the west of the cluster (30Dor-NIC 7, 8) have very
few stars with intrinsic NIR excess. The same is the case around
30Dor-NIC 10, which is located NE of the cluster within the arc,
and NIC 6 and 9, which are located to the south and south-west
of the cluster.

Thirteen of the fourteen fields with broadband observations in all three bands
house at least one source with intrinsic NIR excess (see Tables
\ref{tab1} and \ref{tab2}). Twelve of these fields are shown in Figure
\ref{fig4} with the NIR excess sources identified.

The number of candidate sources identified is
very likely a lower limit to the total number of young stars in the nebular arc
as we required candidates to be detected in all three bands, and as not all
young stellar objects exhibit strong intrinsic IR excess. With six detections,
the 30DOR-NIC12 field exhibits the highest incidence of intrinsic IR excess 
sources. This is probably due to the fact that the location of 30DOR-NIC12 
coincides with one of the density peaks (``pillar'') in the arc.

The comparison with ground-based photometry is restricted to the objects which are in common between our study and the studies by Hyland et al.\ (1992)
and Rubio et al.\ (1998):  the four ``protostars'' P1 to P4
and four objects resolved in knots 1 to 3. The aperture size of 5$''$
as used by Hyland compared to our diffraction limited resolution
of 0\farcs1 (F110W) to 0\farcs2 (F205W) makes a comparison not very meaningful.
The seeing limited observations by Rubio et al.\  (1998) have a better 
resolution
of $\approx$1$''$, but are still by about a factor of 6 to 10 lower spatial
resolution than the HST NICMOS observations. Hence a number of fainter
background sources contribute to the integrated photometry presented
by Rubio et al.\ (1998), which are not included in our photometry due to the 
higher spatial resolution.

A direct comparison to the photometry presented by Rubio et al.\ (1998)
does not yield any evidence for large, systematic differences.
The scatter in the photometric measurements for individual sources
can be explained by the differences in spatial resolution, the non-standard
HST NICMOS passbands (in particular F205W), and last but not least by
intrinsic variability of individual YSOs.

\section{Comparison to previous work on Young Stellar Objects in the 30\,Doradus region}

\subsection{Early O-type stars embedded in nebular knots}

\begin{table*}[htb]
\caption{Near infrared photometry of the brightest individual objects in
knots 1 to 3\label{tab3}}
\begin{tabular}{llccccc} \tableline \tableline
Name&Alias\tablenotemark{1}&$\alpha$(2000) &$\delta$(2000) &J & J--H & H--K$_{\rm s}$\\
       &&[hms]      &[$^\circ$ $'$ $''$] & [mag] & [mag] & [mag]  \\ \tableline
Knot 1a&IRSN-101&5 38 45.37& -69 05 08.9 & 14.67$\pm$0.08 &0.19$\pm$0.09 &0.03$\pm$0.05\\
Knot 1b&IRSN-96&5 38 45.20& -69 05 08.2 & 15.34$\pm$0.06 &0.22$\pm$0.07 &0.20$\pm$0.06\\
Knot 1c&IRSN-96&5 38 45.11& -69 05 08.1 & 16.63$\pm$0.10 &0.38$\pm$0.14 &0.58$\pm$0.19\\ \tableline
Knot 2a&IRSN-135&5 38 48.05& -69 04 43.5 & 14.82$\pm$0.06 &0.31$\pm$0.07 &0.07$\pm$0.04\\
Knot 2b&&5 38 48.00& -69 04 43.5 & 17.01$\pm$0.09 &0.31$\pm$0.12 &-0.14$\pm$0.14\\
Knot 2c&&5 38 47.96& -69 04 43.6 & \nodata &\nodata &\nodata\tablenotemark{2}  \\ \tableline
Knot 3ab\tablenotemark{3} &IRSW-29&5 38 34.55& -69 06 07.2 & 15.61$\pm$0.17 &0.23$\pm$0.24 &0.01$\pm$0.21\\
Knot 3c&&5 38 34.51& -69 06 06.9 & 16.99$\pm$0.05 &1.15$\pm$0.07 &0.28$\pm$0.05\\ \tableline
\end{tabular}

\tablecomments{
$^1$ Rubio et al.\ (1998),
$^2$ m$_{\rm K}$ = 16\fm28$\pm$0\fm08, $^3$ close binary, photometry of
unresolved system, coordinates for center of light}
\end{table*}

Based on optical spectroscopy,
Walborn \& Blades (1987, 1997) identified three early O-type stars 
associated with nebular knots (knots 1 to 3, see Figure \ref{fig4}) in 
the north-east and the
west of 30 Doradus as a first evidence for ongoing star formation in 
the 30 Doradus region. Similar to knot 2, which was originally resolved
into individual IR sources by Rubio et al.\ (1992),  the HST/NICMOS 
observations revealed also knots 1 and 3 as compact multiple systems, very 
much like the
Trapezium system in the Orion cluster (Walborn et al.\ 1999).
Photometry and positions of the brightest components of the multiple
systems are summarized in Table \ref{tab3}. The near infrared colors
corroborate the identification of the subcomponents as early-type stars 
seen through a moderate amount of foreground extinction. 

Assuming that the ratios of projected separations are representative for their
true separations, we find the following ratios (separation of close pair divided
by separation to third component):\\
$\bullet$ Knot 1: 1:2.7 (0.13\,pc vs.\ 0.35\,pc)\\
$\bullet$ Knot 2: 1:1.8 (0.054\,pc vs.\ 0.097\,pc)\\
$\bullet$ Knot 3: 1:2.8 (0.033\,pc vs.\ 0.090\,pc)

For comparison, the four central stars in the Trapezium cluster
in Orion have projected separations between $\approx$0.02\,pc and 
$\approx$0.04\,pc.
Thus similar to the Trapezium system, the three early-type
triple systems in the 30\,Doradus Nebula appear non-hierarchical.
Since non-hierarchical multiple systems are dynamically
unstable (e.g, Mirzoian \& Salukvadze 1985; 
Sterzik \& Durisen 1998; Orlov \& Petrova 2000;
and references therein), the close associations of early-type stars 
in the 30\,Doradus Nebula provides further evidence for their youth. 

As already noted by Walborn et al.\ (1999), the multiple early-type
systems in knots 1 and 2 are associated with a larger number of fainter
stars, and might thus constitute very young open clusters. A third cluster
of faint, extincted stars can be found around 30Dor-NIC15b (see Figure 
\ref{fig2} and \ref{fig4}). 
The on average somewhat larger physical separation of the early type
stars in the center of these open clusters compared to the
Orion cluster might indicate that the former already started
to disperse gradually (see, e.g., Kroupa et al.\ 2001). Alternatively, the 
physical conditions which led
to the formation of the 30\,Doradus young open clusters might have been 
different (less extreme?) than the conditions which led to the birth of the 
Orion cluster. A more detailed discussion of the  stellar populations 
in the association and the young open cluster candidates will be presented
in a forthcoming paper (Grebel et al., in prep.).

\subsection{Evidence for a low-mass cut-off among pre-main sequence
stars in 30 Doradus?}

In their near infrared survey of the 30 Doradus region,
Hyland  et al.\ (1992) found four infrared candidate ``protostars''
with K magnitudes brighter than 12\fm3, and masses in the range of 
15\,M$_\odot$ to 20\,M$_\odot$ (as deduced from the bolometric luminosities
of the infrared sources).
Based on the lack of objects with strong infrared excess and K-magnitudes
between 12\fm3 and 13\fm3 (the latter being the brightness limit of their 
survey), 
Hyland  et al.\ (1992) suggested that there might be a lower-mass cut-off
below 15\,M$_\odot$ in 
the initial mass function for the present-day star formation in 30 Doradus.

The color-magnitude diagram (Figure \ref{fig3}) reveals that the
population of sources with intrinsic IR excess extends down to the 
sensitivity limits of our observations.
The candidate pre-main sequence stars listed in Tables \ref{tab1} \& 
\ref{tab2} have near-infrared colors (and considering the distance to the LMC,
brightness) quite similar 
to Galactic Herbig\,AeBe and T\,Tauri stars with masses in the range from
$\approx$1.5\,M$_\odot$ to $\approx$7\,M$_\odot$. Furthermore, [HJ 91] P1
is resolved into two red objects (30Dor-NIC12b and d, see Figure \ref{fig4}) 
with the fainter one having m$_{\rm K}$ = 13\fm2 (see also Rubio et al.\ 1998).
Thus our HST/NICMOS survey 
gives no evidence for a cut-off in the present-day initial mass 
function possibly down to at least 1.5\,M$_\odot$ 
assuming the infrared excess sources
have properties similar to Galactic pre-main sequence stars.

\subsection{The low-mass IMF in the central starburst cluster}

Sirianni et al.\ (2000) analyzed archived WFPC2 data of the 30\,Dor
region.  The identification of pre-main sequence
stars with masses down to 1.35\,M$_\odot$ close to the center of the
R136 cluster by Sirianni et al.\ (2000) is based on the assumption that 
there is very
little differential reddening towards 30 Doradus. Only then can pre-main
sequence star candidates be identified exclusively by their position in a 
color magnitude diagram. 

Of the NIC2 fields studied by us, only the fields to the west
of R136a overlap with the HST wide field CCD data (though
not the planetary camera data) studied by Sirianni et al.\ (2000).
The NICMOS data clearly reveal the presence of substantial differential
redding in the overlapping fields. Variable extinction
close to the cluster center was reported by Brandl et al.\ (1996)
in their multi-wavelength study of the initial mass function of the
starburst cluster, and is also evident in the data presented by Rubio et al.\ 
(1998) and Scowen et al.\ (1998).  This indicates that a certain fraction 
of the purportedly pre-main-sequence stars in the HST/WFPC2 sample by 
Sirianni et al.\ (2000) might actually be main-sequence stars viewed 
through 5$^{\rm m}$ to 10$^{\rm m}$ of visual extinction. 
An accurate initial mass function can hence only be derived by applying
extinction corrections to individual sources, which in turn requires
multi-color photometry or spectroscopy of individual stars (see also, e.g.,	
Selman et al.\ 1999; Panagia et al.\ 2000).

The scarcity of sources with intrinsic infrared excess inside the
30\,Dor nebular arc, however, does not rule out the presence of more
pre-main sequence stars. The early-type stars in the 30 Doradus cluster
create an intense radiation field, which has an adverse effect on circumstellar
disks. Observations of Galactic HII regions like Orion or NGC 3603 revealed 
that externally illuminated circumstellar disks get photo-evaporated
and disperse on short time scales of typically 10$^4$ to 10$^5$ yr
(e.g., Henney \& O'Dell 1999, Brandner et al.\ 2000a, Richling \& Yorke 2000).

\subsection{Overluminous Herbig AeBe stars?}

Based on data obtained by the EROS project, Lamers et al.\ (1999) identified
a group of irregular variable stars in the bar of the LMC 
with spectral characteristics similar to Galactic Herbig\,AeBe stars.
The majority of these stars, however, appear to be more luminous than Galactic 
Herbig\,AeBe stars, and are located above the birthline (Palla \& Stahler 1993)
of Galactic Herbig\,AeBe stars. Lamers et al.\ (1999) argue that the higher 
luminosity
might be either due to higher overall accretion rates or due to the fact
that pre-main sequence stars in the LMC become optically visible at an
earlier phase than their Galactic equivalents (smaller
dust-to-gas ratio in the LMC). Alternatively, Lamers et al.\ (1999) propose
that they might detect the high end of the luminosity function of 
pre-main sequence stars in the LMC.

In order to test if the infrared luminous ``protostars'' from the work
by Hyland et al.\ (1992) could be representatives of this group
of ``overluminous'' pre-main-sequence stars, we compare their near
infrared photometry to the Galactic Herbig AeBe star Z\,CMa. 
Z\,CMa has a J-magnitude of 5\fm79, a J--K color of 2\fm25,
and is located at a distance of $\approx$1150\,pc (e.g., Leinert et al.\ 1997).
If placed at the distance of the LMC, Z\,CMa would have m$_{\rm J}$=14\fm0
and m$_{\rm K}$=11\fm7, i.e.\ quite in the range of brightness values 
observed for the ``protostars'' from Hyland et al.\ (1992). The redder
J--K colors of the candidate ``protostars'' identified 
by Hyland et al.\ (1992) compared to Z\,CMa could be due to differences
in the geometry of circumstellar material. Edge-on circumstellar disk
sources in Galactic star forming regions can have J--K colors $\ga$7$^{\rm m}$
(e.g.\ Brandner et al.\ 2000b, Table 1).
Thus again we find no strong
evidence that star formation in 30\,Doradus might be different from
Galactic star formation. This is also in agreement with the study of
pre-main sequence candidates around SN\,1987A in 30\,Dor C 
by Panagia et al.\ 2000, 
who do not find evidence for a large population of overluminous
pre-main sequence stars (e.g., Fig.\ 4 in their paper).

These findings do not exclude, however, that star formation in lower
metallicity environments, like the bar region of the LMC,
might be different.

\section{Summary}

We analyzed HST/NICMOS data and groundbased
infrared imaging obtained with the ESO/MPI 2.2m telescope and IRAC2
of the 30 Doradus region.  
We find clear evidence for ongoing, extended star formation in a 10\,pc
$\times$ 15\,pc region
to the north of the central starburst cluster. This region is part of the
arc of molecular gas and warm dust around
30 Doradus. Star formation in this region may have been triggered by the 
starburst cluster.
The fainter IR-excess sources have colors and brightness similar
to Herbig AeBe and T\,Tauri stars with masses possibly in the
range of 1.5\,M$_\odot$ to 7\,M$_\odot$, and could thus be
LMC equivalent of Galactic low- to intermediate-mass
pre-main sequence stars. The present-day initial mass-function apparently
shows no cut-off down to the mass (resp.\ magnitude) limit of our study. The
whole spectrum of stellar masses from pre-main sequence stars with masses
$\ga$1.5\,M$_\odot$ 
to massive O stars still embedded in dense knots might be present in
the nebular filaments.

The  spatial extent of pre-main-sequence stars in the region
to the north of 30 Doradus
suggests that we are witnessing the birth of an OB association including
a small number of open clusters.
Similar older associations, in part with low-density
clusters, are seen throughout the 30 Doradus region
(e.g., Walborn \& Blades 1997; Grebel \& Chu 2000).
Star formation to the west and south appears to be less
intense than in the large scale, nascent OB association to the north of
the cluster. This might be indicative of a different, more
isolated mode of star formation. 
Present day star formation activity in 30\,Doradus coincides with
the spatial distribution of the densest features of the molecular gas,
or their interfaces with the central cavity being evacuated
by the R136 cluster. 

Follow-up near infrared spectroscopy of the pre-main sequence
candidates is required in order to confirm their youth and
to study their physical properties, such as spectral types or
accretion rates. 

\acknowledgements
Support for this work was provided by NASA through grants number
GO-07370.01-96A and GO-07819.01-96A from the Space Telescope Science Institute,
which is operated by the Association of Universities for Research in Astronomy,
Inc., under NASA contract NAS5-26555.
This publication makes use of data products from 2MASS, which is a
joint  project of UMass and IPAC/Caltech, funded by NASA and NSF.
We would like to thank the anonymous referee for her/his insightful
comments, which helped to improve the paper.
WB acknowledges support by NASA and NSF. RHB thanks the Fundaci\'on Antorchas 
(Argentina) for supporting this work.

\end{document}